\documentclass[12pt]{article}
\usepackage{amssymb,amsmath,graphicx}
\begin{document}
\title{\textbf{Gribov copies and anomalous scaling}} 
\author{B. Holdom%
\thanks{bob.holdom@utoronto.ca}\\
\emph{\small Department of Physics, University of Toronto}\\[-1ex]
\emph{\small Toronto, Ontario, Canada M5S1A7}}
\date{}
\maketitle
\begin{abstract}
Nonperturbative and lattice methods indicate that Gribov copies modify the infrared behavior of gauge theories and cause a suppression of gluon propagation. We investigate whether this can be implemented in a modified perturbation theory. The minimal modification proceeds via a nonlocal generalization of the Fadeev-Popov ghost that automatically decouples from physical states. The expected scale invariance of the physics associated with Gribov copies leads to the emergence of a nontrivial infrared fixed point. For a range of a scaling exponent the gauge bosons exhibit unparticlelike behavior in the infrared. The confining regime of interest for QCD requires a larger scaling exponent, but then the severity of ghost dominance upsets naive power counting for the infrared scaling behavior of amplitudes.
\end{abstract}

\section{Introduction}
Gribov \cite{a1} showed that the standard Lorentz covariant gauge fixing procedure based on the Faddeev-Popov (FP) procedure is not complete. Some further modification to the path integral definition of the theory is required to account for the global structure of the field configuration space under gauge transformations. Otherwise gauge equivalent configurations are counted more than once. One source of Gribov copies arises from the existence of different regions of configuration space in which the spectrum of the FP operator has different numbers of negative modes. At a boundary between two such regions an eigenvalue of the FP operator is changing sign. In the first Gribov region the FP operator has no negative eigenvalues, and this region includes the perturbative configurations. The effect of the Gribov boundaries is expected to only become significant at strong coupling in the infrared, where the small eigenvalue modes can play a role.

We consider a $SU(N)$ gauge theory with action $S[A]=\int d^4 x{\cal L}_{YM}$ and ${\cal L}_{YM}=-(\frac{1}{4}F^a_{\mu\nu})^2$. A covariant gauge fixing condition $G(A)=0$ defines a hypersurface in the field configuration space. But a gauge orbit, a set of gauge equivalent configurations, will intersect this hypersurface more than once. To characterize this we consider the following integration along a gauge orbit:
\begin{equation}
1+N(A)=\int{\cal D}\alpha(x)\delta(G(A^\alpha))\left|\det\left(\frac{\delta G(A^{\alpha})}{\delta\alpha}\right)\right|.
\label{e13}\end{equation}
$A^\alpha$ is a gauge transform of $A$ and $\delta G(A^\alpha)/\delta\alpha$ is the FP operator.  $N(A)$ is a measure of the number of Gribov copies, the number of additional intersections of the orbit with the hypersurface, and it is clear that $N(A^\alpha)=N(A)$.\footnote{Note that $N(A)$ does not receive a contribution when the intersection occurs on a Gribov boundary since then the $\det(...)$ vanishes. In such a case the $\delta(...)$ is less constraining and the intersection is expected to consist of a continuous set of copies \cite{a14}.} Of interest also is the fundamental modular region, which is the subregion of the $G(A)=0$ hypersurface such that every gauge orbit intersects it once and only once. It includes the perturbative configurations and it is contained within the first Gribov region, although there is some overlap of the boundaries of the two regions \cite{a13}. 

Even though a description of the fundamental modular region is highly nontrivial, one can use (\ref{e13}) and the insertion of unity trick to at least formally account for the presence of Gribov copies in a path integral \cite{a1},
\begin{equation}
{\cal Z}={\cal N}\int{\cal D}A e^{iS[A]}\delta(G(A))\left|\det\left(\frac{\delta G(A^{\alpha})}{\delta\alpha}\right)\right|\frac{1}{1+N(A)}.
\label{e1}\end{equation}
If we decide to ignore the Gribov copies and as well stay close to the perturbative regime so that the $\det(...)$ does not change sign then we can simplify to the conventional Faddeev-Popov result,
\begin{equation}
{\cal Z}={\cal N}\int{\cal D}A e^{iS[A]}\delta(G(A))\det\left(\frac{\delta G(A^\alpha)}{\delta\alpha}\right).
\label{e9}\end{equation}
The generalized Lorentz gauge condition
$G(A)=\partial^\mu A_\mu^a(x)-\omega^a(x)=0$
with a Gaussian weight for $\omega^a$ leads to the standard gauge fixing addition to the Lagrangian, involving ghost fields $c^a$, $\overline{c}^a$, and a field $B^a$,
\begin{equation}
{\cal L}_{GF}= -\frac{\xi}{2}B^aB^a +B^a (\partial^\mu A_\mu^a)+\overline{c}^a(- \partial^\mu D_\mu^{ac}) c^c.
\label{e15}\end{equation}
The Lagrangian ${\cal L}_{YM}+{\cal L}_{GF}$ has Becchi-Rouet-Stora-Tyutin (BRST) invariance, a global symmetry, which in turn plays an important role in demonstrating perturbative renormalizability and unitarity.

On the other hand this conventional perturbation theory does not offer a complete description in the infrared. If the $\beta$-function of the theory behaves like $\beta(g)\sim -g^\alpha$ for large $g$ and $\alpha>1$ then it suffers from an infrared Landau pole. The running coupling loses meaning below the scale $\Lambda$ where
\begin{equation}
\ln(\frac{\Lambda}{\mu})=\int_{g(\mu)}^\infty \frac{dx}{\beta(x)}.
\label{e14}\end{equation}
When perturbation theory cannot rule out this possibility, for example when there is no Banks-Zaks fixed point \cite{a18}, then any description of the infrared must lie outside of perturbation theory. Normally this is taken to imply something about the nonperturbative structure of the theory, such that the perturbative ambiguities are somehow cancelled or corrected.

But suppose that this breakdown of perturbation theory coincides with the breakdown of the path integral approximation in (\ref{e9}), from which the conventional perturbation theory is derived. This is Gribov's interpretation which links the Landau pole problem to the gauge fixing problem. Gribov \cite{a1} makes the connection by considering the ghost two-point function, which is essentially the inverse of the FP operator in the presence of gauge field fluctuations. Because of the Landau pole, this two-point function also develops a singularity at some spacelike momentum. Thus the FP operator is vanishing, or in other words the gauge fluctuations are driving this operator towards and beyond the first Gribov region. The conventional path integral (\ref{e9}) is thus being pushed beyond its range of applicability. This is an interesting perspective on the Landau pole problem, given the influence the latter has had on investigations into the nonperturbative structure of QCD. 

The form of (\ref{e1}) also suggests that the conventional perturbation theory from (\ref{e9}) is missing a suppression of gauge field fluctuations on scales of order $\Lambda$ and below. First the $|\det(...)|$ is suppressed due to the small eigenvalues in the vicinity of a Gribov boundary. And second, $1/(1+N(A))$ is also likely to cause a relative suppression in the infrared due to the fact that there are gauge copies within and close to the boundary of the first Gribov region, both outside \cite{a13} and on the boundary \cite{a14} of the fundamental modular region. If the copies are occurring closer together and are more numerous then the $1/(1+N(A))$ factor will induce such a suppression.

From the difference between (\ref{e1}) and (\ref{e9}) there is no reason to expect that BRST invariance will be a property of the correct path integral.  In fact it is presently not known how to maintain a local BRST invariance for QCD while accounting for Gribov copies \cite{a12}. The Gribov-Zwanziger approach \cite{a1,a2} at least partially accounts for Gribov copies by constraining the path integral in (\ref{e9}) to the first Gribov region. But this entails a modification to the action that results in the loss of BRST invariance  \cite{a2}, and this remains true in recent extensions of this approach \cite{a21}.

The connection between suppressed gluon propagation in the infrared that Gribov copies seem to imply, and the possible elimination of the Landau pole, has been well discussed in the literature \cite{a8,a10,a4}. Here we shall consider a modification to the action which realizes this connection and which also incorporates the expected scale invariance of the effects of Gribov copies. We are led to consider a nonlocal perturbation theory. One may wonder about the usefulness of this if the modifications are only important in the infrared at strong coupling, but the point is that some statements can be made to all orders in perturbation theory. Depending on the size of a scaling exponent we shall address in this way questions concerning the existence of an infrared fixed point and the validity of naive power counting arguments at such a fixed point. Our findings here should extend beyond the model we study. A result of possible interest for extensions of the standard model is that some gauge theories might exhibit the emergence of unparticlelike \cite{a11} behavior in the infrared.

\section{A Gribov inspired modification}
We propose to model the correct path integral (\ref{e1}) by simply modifying the ghost action of the conventional approach. We shall consider a nonlocal modification that acts to suppress gluon propagation in the infrared while retaining standard perturbation theory to good approximation in the ultraviolet. Landau gauge is often implicitly assumed in the following, as is also the case in the related literature.

First we note that introducing a nonlocality that preserves BRST invariance is not sufficient. Introducing a function $h(\partial^2)$ into the ${\cal L}_{GF}$ in (\ref{e15}),
\begin{equation}
{\cal L}_{GF}^{\rm nonlocal}=-\frac{\xi}{2}B^aB^a +B^a h(\partial^2)(\partial^\mu A_\mu^a)-\overline{c}^a h(\partial^2)\partial^\mu D_\mu^{ac} c^c,
\end{equation}
is just a change in the choice of gauge fixing, $G(A)=h(\partial^2)\partial^\mu A_\mu^a(x)-\omega^a(x)$. This does not eliminate the Gribov copies and the problems of the conventional path integral (\ref{e9}) remain.

We shall consider a modification that does not change the gauge field dependent part of the complete action. The only change is through a modification of ghost propagation, so that the complete Lagrangian is the conventional local one plus a nonlocal term,
\begin{equation}
{\cal L}_{YM}+{\cal L}_{GF}+\overline{c}^a H(\partial^2)\partial^2 c^a.
\end{equation}
The global color symmetry remains while the global BRST symmetry is explicitly broken.

$H(\partial^2)$ must incorporate a scale below which the modification becomes effective, and above which conventional perturbation theory becomes a good approximation. We expect this scale to be similar to the $\Lambda$ defined in (\ref{e14}). We thus explore the simple choice
\begin{equation}
H(\partial^2)=\frac{1}{1+(\frac{\partial^2}{\Lambda^2})^{\kappa}} 
\label{e10}\end{equation}
with $\kappa$ a positive irrational number. In combination with the $-\overline{c}^a\partial^2 c^a$ term in ${\cal L}_{GF}$, the complete ghost kinetic term is suppressed by the factor $(\partial^2/\Lambda^2)^\kappa$ in the deep infrared, leading to an enhanced ghost propagator there. BRST invariance is broken softly.

The infrared enhanced ghost propagator $D^G(p^2)\approx \Lambda^{2\kappa}/(-p^2)^{1+\kappa}$ with $\kappa>0$ has been widely discussed elsewhere. It appears as one of the two conditions in the confinement picture of Kugo-Ojima \cite{a8}. It is required for a well-defined color charge and for the equivalence of BRST-singlet and color singlet states. The other condition is that the gluon propagator must be less singular than a simple pole, which we will see automatically emerges from (\ref{e10}) in our modified perturbation theory. Note that the Kugo-Ojima picture is based on the BRST symmetry of (\ref{e9}) and that the issue of Gribov copies is not addressed.

The behavior $D^G(p^2)\approx \Lambda^{2\kappa}/(-p^2)^{1+\kappa}$ has also been obtained through the nonperturbative study of Schwinger-Dyson (SD) equations \cite{a4} (and also renormalization group methods \cite{a5}), with the further requirement that $\kappa>1/2$. Here the impact of Gribov copies has been discussed by Zwanziger \cite{a10}, who argues that their effect appears only in the choice of boundary conditions for the SD equations. The infrared boundary condition for the ghost self-energy is said to arise from the restriction of the Fadeev-Popov measure to the interior of the first Gribov region (the ``horizon condition''). For the solution of the corresponding SD equation to achieve this behavior, the zeroth order plus perturbative corrections must be cancelled by nonperturbative corrections in such a way that a power-law suppressed behavior for the ghost self-energy can emerge \cite{a10}. This type of cancellation between perturbative and nonperturbative contributions is unusual. And as we have mentioned, BRST symmetry is not available in the Gribov-Zwanziger approach to ensure such a cancellation. Our direct modification of ghost propagation side-steps such cancellations. The results in section 5 will also bring into question the power counting arguments for the scaling behavior of amplitudes as utilized in the SD approach when $\kappa>1/2$.

We also note that a more direct link between Gribov copies and confinement has been discussed \cite{a20}, which does not invoke BRST-symmetry and is in fact quite independent of the form of the action. The ambiguity caused by Gribov copies is shown to directly prevent the construction of physical states with color charge, since the nonlocal dressing function (the analog of the Coulomb tail in QED) required for gauge invariance does not exist in the presence of Gribov copies. The absence of color charges again suggests that QCD differentiates itself from QED via the absence of long-range gauge fields, i.e.~via a gluon propagator that is less singular than a simple pole. In section 4 we shall find that this in itself does not necessarily imply confinement.

The scaling exponent $\kappa$ also determines the power-law suppressed corrections in the ultraviolet. Power-law corrections to perturbation theory are a well-accepted feature of QCD, but they are usually attributed to nonperturbative effects. Here they occur in the modified perturbation theory. These power-law corrections should not affect the perturbative renormalizability of the theory, as expected from the soft breaking of BRST invariance. In particular for irrational $\kappa$ any new ultraviolet divergence governed by $\kappa$ scaling will trivially vanish in dimensional regularization. Note that we are not ruling out a more general choice of $H(\partial^2)$, where the infrared and ultraviolet behaviors are not so simply related.

\section{The ghost and gluon propagators}
Corresponding to the choice in (\ref{e10}) the ghost propagator is
\begin{equation}
D^G(p^2)=\frac{i}{p^2+i\epsilon}-\frac{i\Lambda^{2\kappa}}{(-p^2-i\epsilon)^{1+\kappa}}.
\end{equation}
With $\kappa>0$ this propagator is even more enhanced in the infrared than a standard massless propagator. From this it follows that no K\"allen-Lehmann spectral representation of this propagator exists, which in turn implies that the ghost does not describe a physical asymptotic state. In this sense the modified ghost is unphysical without needing a further subsidiary condition. This is in contrast to the unmodified theory where the ghost does describe an asymptotic state, and it is only after the physical states are required to satisfy the condition $Q_B |\mbox{phys}\rangle=0$, where $Q_B$ is the BRST charge operator, that the ghost along with other unphysical states are removed. Thus we see that $\kappa>0$ appears to be a necessary condition for a unitary theory in the absence of BRST symmetry. Whether or not it is sufficient, and whether the identification of the physical subspace can be accomplished without BRST symmetry, remain open questions.

Although the nonlocal ghosts themselves are unphysical, they can still have an impact on the propagating gluon modes. We have already argued that the effects of a Gribov modification should only start to dominate on scales where the gauge coupling $g$ is large. This reduces the significance of a calculation at leading order in the coupling. Nevertheless it is instructive to consider the one-loop, $\Lambda$ dependent part of the ghost contribution to the gluon vacuum polarization. Using dimensional regularization we find the following.
\begin{eqnarray}
\Pi_{\mu\nu}^\Lambda(p^2)&=&g^2N[c_1\Lambda^{4\kappa}(-p^2)^{-2\kappa}+c_2\Lambda^{2\kappa}(-p^2)^{-\kappa}]g_{\mu\nu}p^2\\&+&g^2N[\tilde{c}_1\Lambda^{4\kappa}(-p^2)^{-2\kappa}+\tilde{c}_2\Lambda^{2\kappa}(-p^2)^{-\kappa}]p_\mu p_\nu\\
c_1&=&\frac{16^\kappa}{512\pi^2}\frac{\cos(\pi\kappa)}{\sin(\pi\kappa)}\frac{1-\kappa}{3-2\kappa}\frac{\Gamma(\kappa-\frac{1}{2})^2}{\Gamma(1+\kappa)^2}\label{e3}\\
c_2&=&\frac{1}{16\pi^2}\frac{1}{\kappa(1-\kappa)(2-\kappa)(3-\kappa)}\label{e3a}\\
\tilde{c}_1&=&2(1-2\kappa)c_1\\
\tilde{c}_2&=&2(1-\kappa)c_2
\end{eqnarray}

$\Pi_{\mu\nu}^\Lambda(p^2)$ is not transverse as expected. We ignore the usual $\Lambda$ independent contributions from ghost and gluon loops which give rise to the running coupling, since we assume that we are in a regime where the power-law dependence in $\Pi_{\mu\nu}^\Lambda(p^2)$ dominates.\footnote{$\Lambda$ must be such that these power-law effects dominate the effects that would generate the Landau pole.} Using  $D_{\mu\nu}(p^2)^{-1}=D_{\mu\nu}^0(p^2)^{-1}-i\Pi_{\mu\nu}^\Lambda(p^2)$ and $D_{\mu\nu}^0(p^2)^{-1}=i(p^2g_{\mu\nu}+(1/\xi-1)p_\mu p_\nu)$, we find by inverting $D_{\mu\nu}(p^2)^{-1}$ and then taking $\xi\rightarrow0$ that
\begin{eqnarray}
D_{\mu\nu}(p^2)&=&\frac{-iZ(p^2)}{p^2+i\epsilon}\left( g_{\mu\nu}-\frac{p_\mu p_\nu}{p^2}\right),\nonumber\\
Z(p^2)^{-1}&=&1-c_1g^2N\Lambda^{4\kappa}(-p^2)^{-2\kappa}-c_2g^2N\Lambda^{2\kappa}(-p^2)^{-\kappa}.
\label{e5}\end{eqnarray}
Thus a transverse propagator is still recovered in Landau gauge. The main point is that the gluon propagator is suppressed in the deep infrared, and thus our modification of ghost propagation has had the desired effect on gluon propagation.
\begin{center}\includegraphics[scale=0.7]{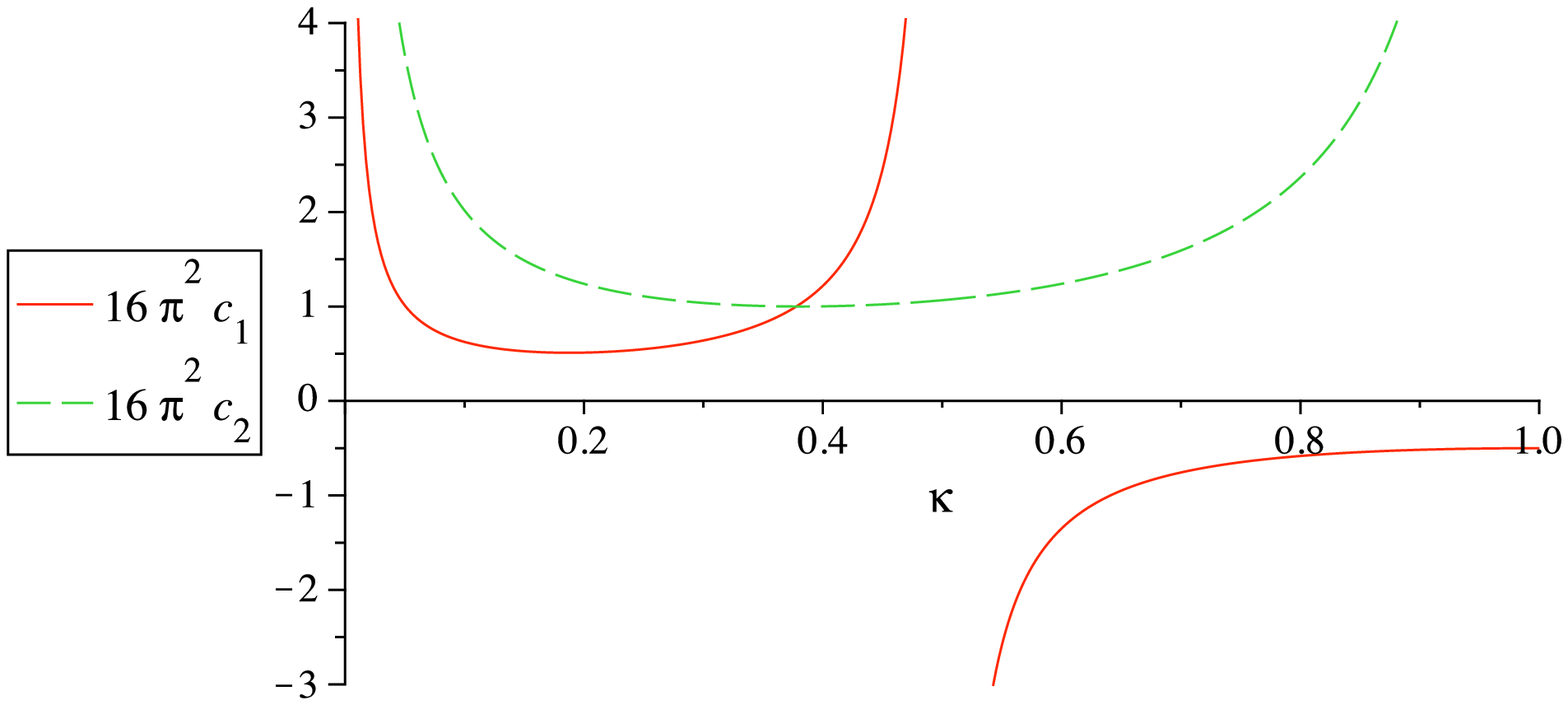}\end{center}
\vspace{-1ex}\noindent Figure 1: $\kappa$ dependence of the leading order coefficients of terms with new power-law behavior in the vacuum polarization (\ref{e5}).
\vspace{2ex}

$ 16\pi^2c_1$ and $ 16\pi^2c_2$ are displayed in Fig.~(1).\footnote{A result equivalent to ours for $c_1$ can be found in the second reference of \cite{a4}.} Because of the behavior of $c_1$ we shall consider separately the two ranges $0<\kappa<1/2$ and $1/2<\kappa<1$. We are supposing that different theories can have different values of $\kappa$, and that it is conceivable for $\kappa$ to fall into either of these two ranges.

\section{The Ungluon}
For the range $0<\kappa<1/2$ both $c_1$ and $c_2$ are positive at one-loop. From (\ref{e5}) it is clear that this would produce a pole on the negative $p^2$ (spacelike) axis. Thus to go any further for this range of $\kappa$ we must assume that the sign of $c_1$ is changed by higher order corrections. If we do this we will have to show that other features of $\Pi_{\mu\nu}^\Lambda(p^2)$ survive such corrections. Assuming $c_1$ is negative then depending on $c_2$ (the value of $c_2$ at all orders) there is a range $0<\kappa<\kappa_{max}$ where possible complex poles have been pushed off the first Riemann sheet. $\kappa_{max}$ ranges from 0 to 1 as $c_2$ ranges from $\sqrt{-4c_1/g^2N}$ down to $-\sqrt{-4c_1/g^2N}$ ($\kappa_{max}=1/2$ when $c_2=0$).

Thus for $0<\kappa<\min(1/2,\kappa_{max})$ the gluon propagator is singular only at $p^2=0$ and has a branch cut along the positive $p^2$ axis. In the deep infrared the propagator approaches the unparticle propagator with an effective gluon scaling dimension of $d_{\cal U}=1+2\kappa$ with $1<d_{\cal U}<2$ \cite{a11}. The `ungluon' is physical in the sense that a positive spectral density function exists. This is contrary to a confining theory such as QCD, where from lattice calculations in Landau gauge \cite{a15} or from analytical arguments \cite{a4} it is found that the gluon propagator exhibits a violation of positivity. Thus to be consistent with confinement we must have $\kappa>1/2$ as we describe in the next section. But it remains possible that a non-QCD-like gauge theory displays the type of behavior discussed in this section, and thus we continue to study this case.

If the theory includes quarks then the ungluon couples to them as a normal gluon. We can write the leading infrared behavior of the current-current amplitude as
\begin{equation}
i{\cal M}\approx\frac{-i}{(\Lambda^2)^{2\kappa}(-p^2)^{1-2\kappa}}\;j_a^\mu j_{a\mu}
\label{e8}\end{equation}
where $j_{a\mu}=\overline{q}\gamma_\mu t_a q$  and $p_\mu$ is the momentum transfer. Thus as $\kappa$ tends towards $1/2$ from below, the current-current interaction tends towards a contact interaction. We do not know the exact normalization, which is $\kappa$ dependent, since we cannot trust the lowest order vacuum polarization. The $p^2$ behavior in (\ref{e8}) is the expected result for an unparticle exchange, although we note how the $(\Lambda^2)^{-2\kappa}$ factor originates in the propagator rather than the vertex. In \cite{a11} the propagator is defined to be independent of $\Lambda$ (the scale below which unparticle behavior occurs) while each unparticle-quark-current vertex has a $\Lambda^{1-d_{\cal U}}=\Lambda^{-2\kappa}$ factor.

We now show how the leading anomalous power-law behavior in $\Pi_{\mu\nu}^\Lambda(p^2)$ does survive to all orders in perturbation theory. We can discuss this along with the behavior of multigluon amplitudes, where invariants constructed from external momenta are characterized by some $P\ll\Lambda$. The dominant one-loop diagram is the ghost loop which is estimated by assuming that the internal momenta are of order $P$. The $m$-gluon amplitude has $m$ internal ghost lines, and thus the truncated amplitude exhibits an enhanced $(\Lambda^2/P^2)^{m\kappa}$ behavior in the infrared. If we consider internal gluon lines as dressed then higher order diagrams can compete with this only if each new gluon line has both ends on ghost lines. Then the number of new ghost lines is always twice the number of new gluon lines, and the enhanced $(\Lambda^2/P^2)^{\kappa}$ behavior of the ghost propagators cancels the suppressed $(\Lambda^2/P^2)^{-2\kappa}$ behavior of the gluon propagators. In other words the original enhancement persists at higher order due to contributions that arise when all internal momenta are characterized by $P$. The dominant diagrams can contain any number of ghost loops, but no other elementary vertex besides the gluon-ghost vertex can appear.

We can extend the argument to the unphysical amplitudes involving external ghosts, with or without external gluons. Here the tree level diagram contributes to the leading infrared behavior. The higher order contributions that can compete again involve ghost loops with gluon lines that only end on ghosts. From any such diagram we find that the infrared behavior for truncated amplitudes with $m$-gluons and $2n$-ghosts has the anomalous scaling factor
\begin{equation}
\Gamma^{m,n}\sim(\Lambda^2/P^2)^{(m-n)\kappa}.
\label{e2}\end{equation}
The behavior of the vacuum polarization ($m=2$, $n=0$) and the ghost self-energy ($m=0$, $n=1$) shows that the power counting has been self-consistent. The signs of amplitudes are left undetermined by these arguments, and we have already made use of this to choose the sign of $c_1$. The scaling in (\ref{e2}) reproduces and is occurring in a way similar to that obtained from Schwinger-Dyson equations \cite{a4}, although at this stage we are only considering the case $0<\kappa<1/2$.

Effective couplings among gluons and ghosts are determined by the truncated amplitudes times appropriate factors of the wave function renormalizations. These latter factors are $(\Lambda^2/P^2)^{-\kappa}$ and $(\Lambda^2/P^2)^{\kappa/2}$ for each external gluon and ghost line respectively. These factors cancel the one in (\ref{e2}), and thus the effective dimensionless couplings approach constants. The theory approaches a nontrivial infrared fixed point \cite{a4}.

If external quark lines are considered then the truncated amplitudes are suppressed by $(\Lambda^2/P^2)^{-n_q\kappa}$ for $2n_q$ external quark lines.\footnote{Diagrams with internal quark lines are suppressed further.} In this case the quark wave function renormalizations do not supply the compensating enhancements, and thus the effective couplings among quarks are suppressed in the infrared [as in (\ref{e8}) for example]. Quarks behave as normal particles that interact through unparticle exchange.

It is interesting to consider the ``deconstruction'' of the ungluon into a tower of massive gluons \cite{a6}, and then consider the longitudinal degrees of freedom of these massive gluons. Instead of arising from Goldstone bosons, the longitudinal modes are due to dressed ghost-antighost pairs. There is some similarity to the massless Schwinger model in two dimensions, where the photon mass arises from a fermion loop. There the fermion loop produces a pole rather than a branch cut, and this in turn implies that the massive boson does not decay into the massless fermions. The massless fermions do not exist in the physical spectrum, a result that does not depend on a subsidiary condition. Our case is similar, with the normal branch cut now being replaced by an unparticle branch cut. Correspondingly the tower of massive gluons are stable; they do not decay to ghosts. This is consistent with the nonlocal ghosts not appearing in the physical spectrum, without need of a subsidiary BRST condition as already noted.

In summary, a Gribov inspired modification leads to the emergence of a nontrivial infrared fixed point in gauge theories that would otherwise suffer from an infrared Landau pole. This fixed point is accessible in a perturbation theory analysis. The gauge fields may emerge as unparticles and we have noticed that the physics of unparticles helps to make sense of the lack of BRST invariance. There is a superficial resemblance to the confinement picture of Kugo-Ojima \cite{a8}, since the gluon and ghost propagators display similar infrared behavior. But the unparticle picture that emerges in our case indicates that confinement is not occurring. Rather the quarks interact in the infrared through unparticle exchange (\ref{e8}), with this interaction becoming arbitrarily weak compared to massless gluon exchange in the infrared limit.

\section{Ghost dominance}
For the range $1/2<\kappa<1$ the one-loop vacuum polarization gives $c_1<0$ and $c_2>0$. This produces a pair of complex conjugate poles in the timelike half-plane if $c_2^2<|4c_1/g^2N|$. This will occur for $\kappa$ in some range $1/2<\kappa<\kappa^0_{max}<1$ from Fig.~(1), if we continue to take the one-loop results seriously. The interpretation of such poles is uncertain, but they have been discussed within a confining picture in \cite{a7}. On the other hand if we assume that higher order corrections produce a negative $c_2$ as well, then as in our discussion at the beginning of the last section there is a range $1/2<\kappa<\kappa_{max}<1$ where the complex poles have been pushed off the first Riemann sheet.

In any case the effective scaling dimension of the gluon in the deep infrared is now $2<d_{\cal U}<3$. This takes us outside the range for an unparticle, but it can be considered here because we have an ultraviolet completion. The gluon propagator again has a branch cut along the positive $p^2$ axis, but now the propagator vanishes for $p^2=0$. This vanishing is also a property of the Gribov-Zwanziger approach \cite{a1,a2}, and it means that there must be both positive and negative norm contributions in the spectral density function. This violation of positivity is expected for confinement as we have already mentioned. 

But it is overly simplistic to extend the infrared scaling behavior of amplitudes in the last section to the present case. In particular it was argued that the diagrams were dominated by internal momenta of order the small external momenta $P$, but this may not be the case for the highly suppressed internal gluon lines when $\kappa>1/2$. For example the single dressed gluon exchange between quark currents as indicated by (\ref{e8}) would vanish as the momentum transfer vanishes. Clearly there are larger contributions from physics at the $\Lambda$ scale, for example from multiple gluon exchange. Integrating out the dressed gluons will give rise to effective operators such as $j^\mu j_{\mu}$ and $j_a^\mu j_{a\mu}$, where $j_{\mu}=\overline{q}\gamma_\mu q$ and $j_{a\mu}=\overline{q}\gamma_\mu t_a q$. For momentum transfers much below $\Lambda$ these induced effectively local interactions will have strength of order $1/\Lambda^2$. The typical internal momenta are of order $\Lambda$ since the momentum integrations are convergent in both the infrared and ultraviolet.

But we can identify still larger contributions. Integrating out the gluons will also generate effective interactions between quarks and ghosts, such as $j^{\mu}j^{\rm gh}_\mu/\Lambda_1^2$ and $j_a^\mu j^{\rm gh}_{a\mu}/\Lambda_2^2$ where $j^{\rm gh}_\mu=\overline{c}_b \partial_\mu c_b$, $j^{\rm gh}_{a\mu}=\overline{c}_b f^{bac} \partial_\mu c_c$ and $\Lambda_1\approx\Lambda_2\approx\Lambda$. Now consider the interactions between quark currents that are induced by these effective operators and a ghost loop, as shown in Fig.~(2a). When $\kappa>1/2$ the ghost loop integral is dominated in the infrared, such that a momentum transfer small compared to $\Lambda$ acts as the infrared cutoff. Then the ultraviolet cutoff implied by the effective operators can be ignored, and our previous one-loop vacuum polarization calculation is easily adapted. The induced current-current amplitudes are
\begin{eqnarray}
i{\cal M}&=&\frac{i{c}_1(N^2-1)}{\Lambda_1^4(\Lambda^2)^{-2\kappa}(-p^2)^{2\kappa-1}}\;j^\mu j_{\mu}\nonumber\\&-&\frac{i{c}_1N}{\Lambda_2^4(\Lambda^2)^{-2\kappa}(-p^2)^{2\kappa-1}}\;j_a^\mu j_{a\mu}
\label{e4}\end{eqnarray}
where $c_1$ is given in (\ref{e3}).

Thus as $\kappa$ increases from $1/2$ the current-current interactions in the infrared move away from the contact form and grow stronger. The trend observed in the last section is reversed. We note again that the ghost loop does not generate a normal two-particle cut, and in fact these current-current interactions again look like unparticle exchange with $1<d_{\cal U}<2$. We also comment on the signs in (\ref{e4}); since $c_1$ is negative we have attractive (repulsive) interactions between quarks and antiquarks from the noncolored (colored) currents respectively.

But this is not the end of the story. We may also expect that effective four-ghost interactions arise when integrating out the gluons. This will completely change the power counting of the last section for the gluon plus ghost sector of the theory. In other words there are more important contributions to amplitudes arising when subdiagrams in a diagram are characterized by momenta of order $\Lambda$ rather than $P$. These subdiagrams are being represented by effective operators, and for now we consider the four-ghost operators $j^{{\rm gh}\mu}j^{\rm gh}_\mu/\Lambda_3^2$ and $j_a^{{\rm gh}\mu} j^{\rm gh}_{a\mu}/\Lambda_4^2$.

\vspace{1ex}\begin{center}\includegraphics[scale=0.5]{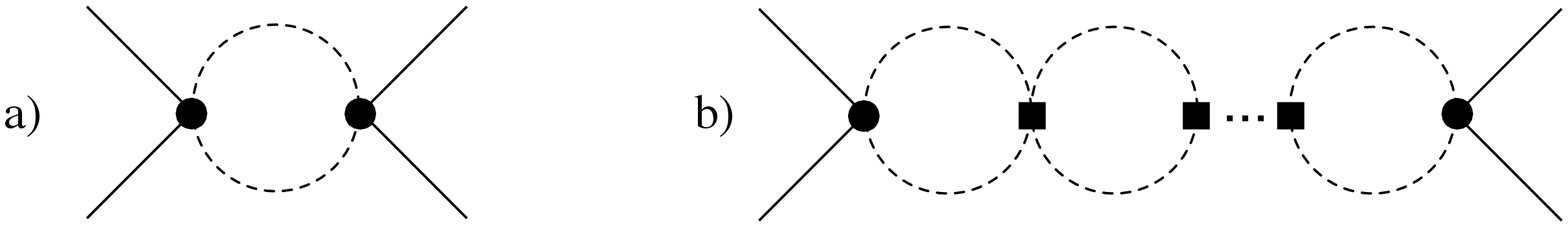}\end{center}
\vspace{-1ex}\noindent Figure 2: Ghost loop contributions to the scattering of quarks
\vspace{2ex}

For example we see that $n$ insertions of the appropriate four-ghost operator can transform a ghost loop into a chain of ghost loops (Fig.~(2b)). Now suppose these chains are naively summed up. Then the amplitudes in (\ref{e4}) are replaced by
\begin{eqnarray}
i{\cal M}&=&\frac{i{c}_1(N^2-1)}{\Lambda_1^4(\Lambda^2)^{-2\kappa}(-p^2)^{2\kappa-1}}\left[\frac{1}{1-\eta_3{c}_1N^2[\Lambda^2/\Lambda_3^2][\Lambda^2/(-p^2)]^{2\kappa-1}}\right]\;j^\mu j_{\mu}\nonumber\\&-&\frac{i{c}_1N}{\Lambda_2^4(\Lambda^2)^{-2\kappa}(-p^2)^{2\kappa-1}}\left[\frac{1}{1-\eta_4{c}_1N[\Lambda^2/\Lambda_4^2][\Lambda^2/(-p^2)]^{2\kappa-1}}\right]\;j_a^\mu j_{a\mu}.
\label{e12}\end{eqnarray}
$\eta_3=\pm1$ and $\eta_4=\pm1$, depending on the sign of the respective four-ghost operator. 

First consider the $j_a^\mu j_{a\mu}$ amplitude when $\eta_4=+1$, where we see that it tends smoothly to a constant in the deep infrared. In fact our previous calculation of the vacuum polarization $\Pi_{\mu\nu}^\Lambda(p^2)$ is also modified by these same chains of ghost loops.\footnote{We could have specified from the beginning that all effective operators exclude the one gluon contribution, thus making the chains 1PI.} The net effect is to multiply the $c_1$ term in (\ref{e5}) by the square bracket multiplying $j_a^\mu j_{a\mu}$ in (\ref{e12}). This modification only applies in the infrared. The implication is that the gluon propagator itself now tends to a constant in the deep infrared. This argument survives corrections to the basic ghost loop, since any new behavior of the corrected loop will cancel out after summing the chains.

Now let us briefly consider the $j^\mu j_{\mu}$ amplitude in (\ref{e12}) when $\eta_3=-1$. This yields a spacelike singularity in the scattering of quarks in the color singlet exchange channel. In this case we appear to be recovering a form of infrared slavery, generated by the exchange of self-interacting ghost-antighost pairs rather than gluons. This may illustrate a more generic infrared instability in the perturbative expansion of scattering amplitudes, since the insertion of more four-ghost operators into diagrams tends to cause further infrared enhancement. The suggestion here is that the physical states of the theory (the color singlets?) are those that do not experience these ghost mediated interactions.

\section{Summary}
We have investigated some consequences of having nonlocality creep into gauge theories ``through the back door'', via the covariant gauge fixing sector and its inherent Gribov copies. We have considered a minimal modification to perturbation theory which enhances ghost propagation in the infrared. This enhancement in turn causes suppression of gluon propagation, while at the same time ensuring that the ghost decouples from the physical sector. The expected scale invariance of the effects of Gribov copies shows up as a nontrivial infrared fixed point.

In section 5 we described how the very enhanced nature of ghost propagation when $\kappa>1/2$ can be responsible for nontrivial long-range interactions among quarks. A related picture is described in \cite{a10}. We have highlighted a separation between the characteristic scale $\Lambda$ of gluon propagation, which induces the ghost-quark interaction, and the infrared dominance of ghost propagation. The result is the loss of the simple scaling behavior of general amplitudes described in section 4, which are thus only true when $\kappa<1/2$. We note that the study of Schwinger-Dyson equations in Landau gauge indicate that $\kappa>1/2$ for QCD \cite{a4}. That work invokes power counting arguments similar to section 4 for the scaling of general amplitudes, but we are claiming that this is no longer correct for $\kappa>1/2$.  This observation would seem to have important implications for the SD approach. In particular we saw how this could cause the gluon propagator to approach a constant in the infrared rather than vanish, and this is of interest in light of recent lattice results \cite{a16}.

Our results for $\kappa<1/2$ show that this case cannot describe a confining theory. Instead the gauge field displays unparticlelike behavior as the theory approaches a nontrivial infrared fixed point. Such a gauge theory could still exhibit chiral symmetry breaking, and would thus be of interest for electroweak symmetry breaking. Heavy fermions might then strongly experience the unparticle interaction. In this context it would also be interesting to know whether the unparticlelike behavior could survive the breaking of the remaining global color symmetry. If so, then gauge fields with unparticle behavior may also be of interest for the breaking of gauged flavor symmetries, and this leads to the speculation that light fermions could experience very weak unparticle interactions as remnants of flavor symmetries.

\section*{Acknowledgments}
This work was supported in part by the National Science and Engineering Research Council of Canada.

\end{document}